\documentclass[final,5p,times,twocolumn]{elsarticle}

\usepackage{graphicx}
\usepackage[dvips]{color}
\usepackage[figuresright]{rotating}

\usepackage{amssymb}
\usepackage{amsmath}

\usepackage{url}

\journal{Nucl. Instr. and Meth. in Phys. Res. A}

\begin{document}

\begin{frontmatter}



\title{Characterization and Simulation of the Response of Multi Pixel
  Photon Counters to Low Light Levels }

%% Authors and addresses
\author[ic]{A.~Vacheret}
\author[war]{G.J.~Barker}
\author[warsaw]{M.~Dziewiecki}
\author[ic]{P.~Guzowski}
\author[war,oxf]{M.D.~Haigh}
\author[lsu]{B.~Hartfiel}
\author[INR]{A.~Izmaylov}
\author[csu]{W.~Johnston}
\author[INR]{M.~Khabibullin}
\author[INR]{A.~Khotjantsev}
\author[INR]{Yu.~Kudenko}
\author[warsaw]{R.~Kurjata}
\author[lsu]{T.~Kutter}
\author[ubc]{T.~Lindner}
\author[ic]{P.~Masliah}
\author[warsaw]{J.~Marzec}
\author[INR]{O.~Mineev}
\author[INR]{Yu.~Musienko}
\author[ubc]{S.~Oser}
\author[triumf]{F.~Reti\`ere\corref{cor1}}
\ead{fretiere@triumf.ca}
\author[shef]{R.O.~Salih}
\author[INR]{A.~Shaikhiev}
\author[shef]{L.F.~Thompson}
\author[shef]{M.A.~Ward}
\author[csu]{R.J.~Wilson}
\author[INR]{N.~Yershov}
\author[warsaw]{K.~Zaremba}
\author[warsaw]{M.~Ziembicki}

\cortext[cor1]{Corresponding author}

\address[ubc]{Department of Physics \& Astronomy, University of British Columbia,  6224 Agricultural Road, Vancouver, BC V6T 1Z1, Canada}
\address[csu]{Department of Physics, Colorado State University, Fort Collins, CO 80523, USA}
\address[ic]{Department of Physics, Imperial College London, South Kensington Campus, London SW7 2AZ, UK}
\address[INR]{Institute for Nuclear Research RAS, 60 October Revolution Pr. 7A, 117312 Moscow, Russia}
\address[lsu]{Department of Physics and Astronomy, Louisiana State University, 202 Nicholson Hall, Tower Drive, Baton Rouge, Louisiana 70803, USA}
\address[triumf]{TRIUMF, 4004 Wesbrook Mall, Vancouver, BC V6T 2A3, Canada}
\address[shef]{Department of Physics and Astronomy, University of Sheffield, Hicks Building, Hounsfield Rd. Sheffield S3 7RH, UK}
\address[war]{Department of Physics, University of Warwick, Gibbet Hill Road, Coventry CV4 7AL, UK}
\address[warsaw]{Institute of Radioelectronics, Warsaw University of Technology, 15/19 Nowowiejska St., 00-665 Warsaw, Poland}
\address[oxf]{Current address: Department of Physics, The University of Oxford, Clarendon Laboratory OX1 3PU, Oxford, UK}

\begin{abstract}
The calorimeter, range detector and active target elements of the T2K near detectors
rely on the Hamamatsu Photonics Multi-Pixel Photon Counters (MPPCs) to detect
scintillation light produced by charged particles. Detailed
measurements of the MPPC gain, afterpulsing, crosstalk, dark noise,
and photon detection efficiency for low light levels are reported.
In order to account for the impact of the MPPC behavior on T2K physics observables, a
simulation program has been developed based on these measurements.
The simulation is used to predict the energy resolution of the detector.

\end{abstract}

\begin{keyword}
photosensors \sep photodetectors \sep multi-pixel avalanche photodiodes \sep scintillator
\PACS 29.40.Mc \sep 29.40.Wk \sep 29.40.Wj
\end{keyword}

\end{frontmatter}


\section{Introduction}

The Tokai to Kamioka (T2K) project~\cite{t2k} is a second-generation
long-baseline neutrino oscillation experiment that uses a high
intensity off--axis neutrino beam produced by the 30~GeV proton beam
at the Japan Proton Accelerator Research Complex (J-PARC).  The first
phase of the T2K experiment pursues two main goals: a sensitive
measurement of ${\rm\theta}_{13}$, and determination of the parameters
${\rm sin^22\theta}_{23}$ and $\Delta m^2_{23}$ to better accuracy
than any previous experiment.

\begin{figure*}[htb]
\centering\includegraphics[width=0.8\textwidth]{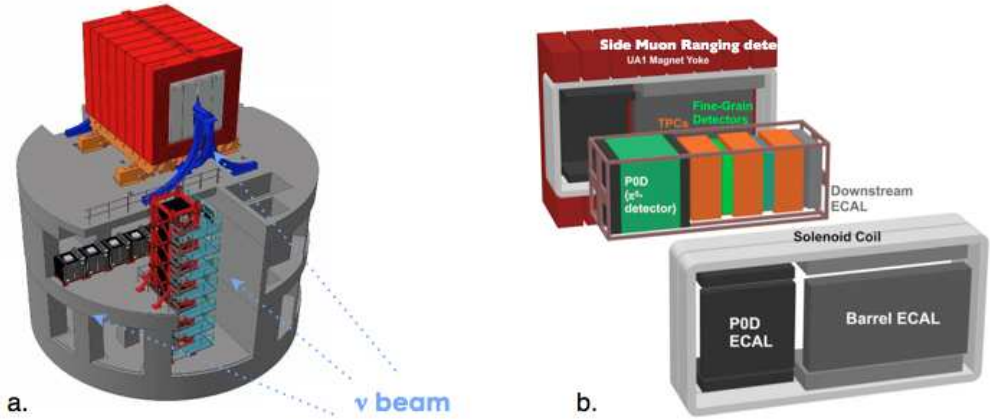}
\caption{Schematic view of (a) the T2K ND280 near detector complex consisting of the on--axis neutrino beam monitor (the ``cross'' configuration of cubical black modules on the two lower levels) and off--axis near neutrino detector on the top level, and (b) an exploded view of the off--axis near neutrino detector.}
\label{fig:nd280}
\end{figure*}

To reach these physics goals, precise knowledge of
the neutrino beam flux and spectrum, and the neutrino
interaction cross sections is required. To perform the required measurements, the near detector complex (ND280~\cite{nd280}) was built at a distance of 280~m from the
hadron production target. The complex has two detectors
(Fig.~\ref{fig:nd280}): an on--axis detector (neutrino beam monitor),
and an off--axis neutrino detector located along the line between the
average pion decay point and the Super-Kamiokande detector, at
2.5$^{\circ}$ relative to the proton beam direction.  The on--axis
detector (INGRID) consists of $7 + 7$ identical modules, arranged to form a
``cross'' configuration, and two ``diagonal'' modules positioned off the cross axes.
The off--axis detector includes a magnet, previously used in the UA1 and NOMAD experiments, operated with a magnetic field of up to 0.2~T; a Pi-Zero detector (POD); a tracking detector that includes time projection chambers (TPCs) and fine
grained scintillator detectors (FGDs); an electromagnetic calorimeter
(ECAL); and a side muon range detector (SMRD).

The ND280 detector extensively uses scintillator detectors and embedded
wavelength-shifting (WLS) fibers, with light detection from the
fibers by photosensors that must operate in a magnetic field and fit in limited space inside the magnet.

After studying several candidate photosensors, a multi-pixel avalanche photodiode
operating in the limited Geiger multiplication mode was selected as
the photosensor. These novel devices are compact, well
matched to the spectral emission of WLS fibers, and insensitive to
magnetic fields.  Detailed information about such devices and basic
principles of operation can be found in recent review papers (see for
example \cite{renker} and references therein).

The operational parameters required for these photosensors were similar
for all ND280 subdetectors and can be summarized as follows: an active
area diameter of $\sim$1~mm$^2$, photon detection efficiency for
green light $\geq$20\%, a gain of $(0.5-1.0)\times10^6$, more than 400
pixels, and a single photoelectron dark rate $\leq1$~MHz. The pulse width should be
less than 100~ns to match the spill structure of the J-PARC proton
beam.  For calibration and control purposes it was very desirable to
obtain well-separated single electron peaks in the amplitude spectra for
dark noise and low light levels.

\begin{figure}[htb]
\centering\includegraphics[width=0.5\textwidth]{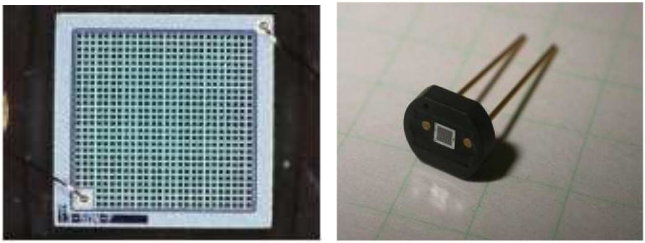}
\caption{Photographs of an MPPC with a sensitive area of $1.3\times
  1.3$ mm$^2$: magnified face view (left) with 667 pixels in a $26\times
  26$ array (9 pixels in the corner are occupied by an electrode); the ceramic
  package of this MPPC (right).}
\label{fig:mppc}
\end{figure}

After an R\&D study period of three years by numerous groups, the
Hamamatsu Multi-Pixel Photon Counter (MPPC) was chosen as the
photosensor for ND280. A description of this type of device and its
basic parameters can be found in Ref.~\cite{mppcSpec}. A customized
667-pixel MPPC with a sensitive area of 1.3$\times$1.3~mm$^2$ was
developed for T2K~\cite{mppcT2K}. It is based on the Hamamatsu commercial
device S10362-11-050C with 400 pixels and 1$\times$1~mm$^2$ sensitive
area. The sensitive area was increased to provide better acceptance of
light from 1~mm diameter Y11 Kuraray fibers. In total, about 60,000
MPPCs were produced for T2K. The sensor is shown in
Fig.~\ref{fig:mppc}.

In this paper, we present the results of measurements and simulations
of the main parameters of Hamamatsu MPPCs developed for the T2K
experiment, expanding upon the results given in \cite{mppcChar}.
We emphasize the operational parameters of these devices
most critical for successful operation and calibration of the T2K
ND280 detectors: gain, dark rate, crosstalk, afterpulses and photon detection efficiency. This paper complements the results reported
in~\cite{mppcQA}, which focused on assessing the gross features of a
large number of MPPCs. In this paper, dedicated setups were built to
measure each process, which enabled more in-depth measurements
than in~\cite{mppcQA} but in general these setups did not allow testing
of a large number of MPPCs.

\section{MPPC response}\label{sec_lowlight}
%-----------------------------------------------------------
\subsection{Operating principles}
%-----------------------------------------------------------

A Multi-Pixel Photon Counter consists of an array of avalanche
photo-diodes operating in Geiger mode. When operating in Geiger mode
the diode is reverse-biased beyond the electrical breakdown voltage,
which will be denoted $V_{\rm BD}$ throughout this document. Above
$V_{\rm BD}$, the electric field in the diode depletion region is
sufficiently large for free carriers to produce additional carriers by impact
ionization, resulting in a self-sustaining avalanche. In practice
irreversible damage would eventually occur unless the avalanche is
quenched. In MPPCs, quenching is achieved by using a large resistor in
series with the diode. The current produced by the avalanche creates a
voltage drop across the resistor ($R_{\rm quench}$), which stops the
avalanche when the voltage across the diode reaches $V_{\rm BD}$. The
overvoltage, denoted $\Delta V$, is the difference between the
operating voltage of the device and the breakdown voltage $V_{\rm BD}$.
The charge produced in an avalanche is hence the diode capacitance
times $\Delta V$.

In Geiger mode, the amount of charge produced in an avalanche is
independent of the number of charge carriers generated within the
depletion region. Hence, it is not possible to measure the light
intensity by measuring the total charge produced in a single
avalanche. MPPCs achieve photon counting capability by segmenting
the detection area in an array of
individual diode pixels. The amount of light hitting the device is
sampled by counting the number of pixels that produce avalanches, which leads
to a saturation effect when a large amount of light hits the
sensor. However, the focus of this paper is the MPPC response to low light
levels, where the probability that multiple photons hit the same pixel at the same time is small.

The T2K MPPC is an array of 26 by 26 pixels, each of which measures $50\times50~\mu$m$^2$, on a common n$++$--type silicon substrate~\cite{yamamoto}. Nine
pixels in one corner have been replaced by a lead, reducing the total
number of pixels to 667. The quenching resistors are
polysilicon resistors. The Hamamatsu specifications sheet~\cite{mppcSpec} states
that the fill factor, i.e. the fraction of the device area that is
active, is 61.5\%. The breakdown voltage is about 70~V. When devices are purchased from Hamamatsu, rather than providing the breakdown voltage for each device, the voltage necessary to achieve a gain of $7.5\times10^{5}$ at $25^\circ$C is provided.

%-----------------------------------------------------------
\subsection{Electrical properties}\label{elec_prop}
%-----------------------------------------------------------

\begin{figure}[htb]
\centering\includegraphics[width=0.5\textwidth]{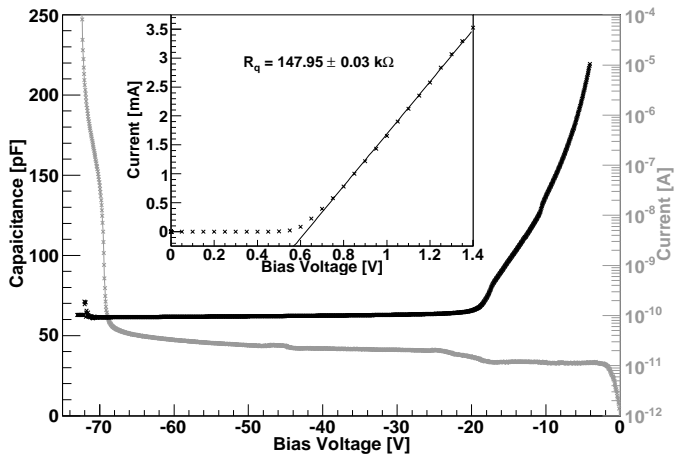}
\caption{$I-V$ and $C-V$ plots for an MPPC.}
\label{fig:cvv}
\end{figure}

The total resistance and capacitance of an MPPC were measured using a
picoammeter and capacitance-voltage (CV) analyzer,
respectively. $I-V$ and $C-V$ plots are shown in
Fig.~\ref{fig:cvv}. The MPPC capacitance was measured with a
Keithley~590 CV analyzer. The capacitance drops rapidly with voltage
down to -20~V, which presumably corresponds to the full depletion of
the device. The capacitance of the MPPC was found to follow a linear
relationship when the supply voltage is less than -20~V:
$C_{\rm MPPC}=aV+b$ with $a$=0.0436$\pm$0.0003~pF/V and $b$=64.27$\pm$0.01~pF.
At -70~V, the capacitance is then 61.22$\pm$0.02~pF. The
Hamamatsu specification document for T2K's MPPCs states that the
terminal capacitance is 60~pF, which is consistent with 61.28~pF
obtained at -70~V operating voltage. In the remainder of this paper, the minus
sign will be omitted when discussing operating voltage. Using 60~pF
total capacitance and neglecting parasitic capacitance yields a pixel
capacitance of $C_{pix}=90.0$~fF.

The current was measured with a Keithley 617 programmable electrometer
at 23$^{\circ}$C. A linear fit for a forward bias voltage larger than
0.6~V yields a slope of $R_{\rm quench}/(667~{\rm pixels})$=225~$\Omega$.
From this we determine the average quenching resistor value for this device to be $R_{\rm quench}=150~{\rm k}\Omega$; for a set of thirty five sensors this parameter was distributed in the range 148--154~k$\Omega$.

%-----------------------------------------------------------
\subsection{Recovery time}\label{recovery}
%-----------------------------------------------------------
When an avalanche occurs in a pixel, the bias voltage across the diode
drops down to the breakdown voltage. The diode voltage recovers to the
nominal operating voltage with a time constant that is nominally given
by the product of the pixel capacitance and the quenching
resistor. Using the values of $R_{\rm quench}$ and $C_{\rm pixel}$ reported in
the previous section, the recovery time constant is $\tau$=13.4~ns. The overvoltage on the pixel at time $t$ after the avalanche
can then be written as: $\Delta V(t)=\Delta V(0) (1-e^{-t/\tau})$,
where $\Delta V(0)$ is the nominal overvoltage. We will see in the
following section that the MPPC behavior is almost entirely driven by
the overvoltage. Lower overvoltage implies a lower probability of
triggering an avalanche. It also implies a lower MPPC gain, hence an
avalanche occurring while the pixel is recovering will yield a lower
charge.

The pixel voltage recovers to its nominal value by pumping charge from
neighboring pixels and from the external electronics circuit. The
capacitance of one pixel (90~fF) is small compared to the total
capacitance of the MPPC (60~pF). Hence the voltage drop induced by the
avalanche in one pixel on all the other pixels is very
small. However, the neighboring pixels effectively act as a bypass
capacitor and the external circuit must eventually recharge the whole
MPPC. The time constant introduced by the external circuit may be much
longer than the pixel $RC$ time constant and should be taken into account when
investigating the response of the MPPC to large light pulses, or when
the repetition rate of avalanches is high.
Since here we focus on characterizing the MPPC
response to low light levels ($<$100 photoelectrons), the impact of the
external electronics on the recovery time can be neglected.

%-----------------------------------------------------------
\subsection{Photosensor gain}\label{gain}
%-----------------------------------------------------------

\begin{figure}[htb]
\centering\includegraphics[width=0.5\textwidth]{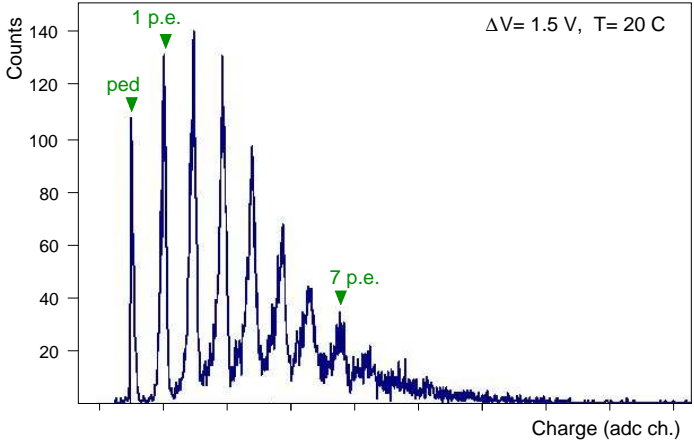}
\caption{A charge amplitude spectrum obtained using an LED source
measured with an MPPC (serial number TA9445) at an overvoltage of
1.5~V and temperature of 20$^{\circ}$C.}
\label{fig:adc_spectrum}
\end{figure}

The MPPC gain is defined as the charge produced in a single pixel
avalanche, expressed in electron charge units. Single avalanches are
typically created by a single carrier (unit charge) and can be
triggered either by a photon or by thermal
noise. Fig.~\ref{fig:adc_spectrum} demonstrates excellent separation
between the charges resulting from different number of
photoelectrons. The gain is measured using a Multi-Channel Analyzer
(MCA), by taking an amplitude spectrum and calculating the distance
between the pedestal peak and the single photoelectron peak. Using
other peaks provides consistent results. Conversion from the MCA
output in digital counts to units of charge is achieved by calibrating
the electronics with a known input charge.
The accuracy of the absolute gain measurement (i.e. the charge corresponding to a single
avalanche) is affected mainly by the accuracy of the charge injection calibration.

\begin{figure}[htb]
\centering\includegraphics[width=0.5\textwidth]{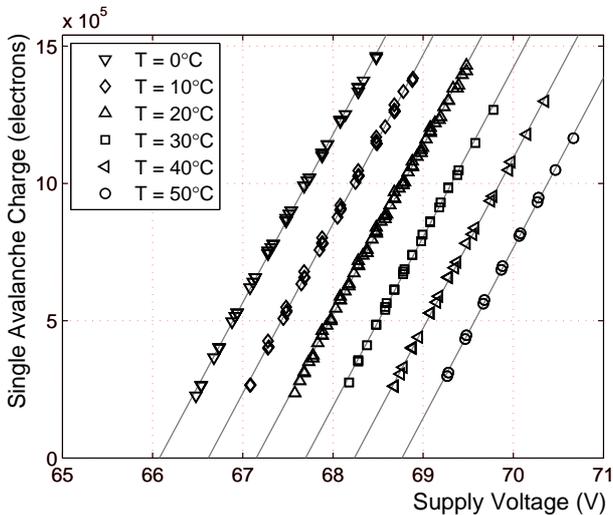}
\caption{MPPC gain vs. supply (bias) voltage at different temperatures
(sensor serial number TA8120).}
\label{fig:gain}
\end{figure}

Fig.~\ref{fig:gain} shows the gain as a function of operating voltage
for various temperatures. The curves are fit by linear functions
according to gain, $G=C_{\rm pix}(V-V_{\rm BD})$, where $C_{\rm pix}$ denotes the single
pixel capacitance, $V$ the operating voltage, and $V_{\rm BD}$ the
breakdown voltage, i.e. the voltage for which extrapolation of the curves produces zero
gain. The curves exhibit a slightly quadratic dependence, but a linear fit gives a reasonable estimate of $V_{\rm BD}$ and will be used throughout this paper.
[We note that the voltage dependence of $C_{\rm MPPC}$ reported in Section~\ref{elec_prop} would cause the gain to have a quadratic dependence but this effect is smaller than the quadratic dependence we observe.]
Since $V_{\rm BD}$ increases linearly by 52$\pm4$~mV per $^{\circ}$C, the gain decreases proportionately as the temperature increases at fixed operating voltage. However, the temperature variations within the T2K ND280 experiment are small enough that this effect can be calibrated out and does not require active compensation.

\begin{figure}[htb]
\centering\includegraphics[width=0.5\textwidth]{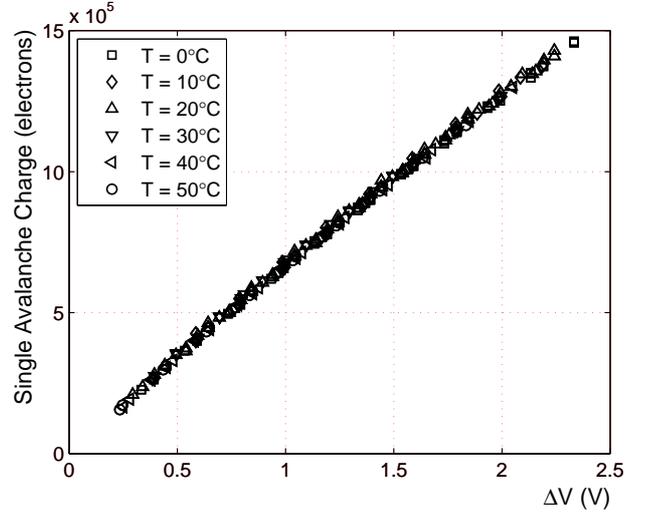}
\caption{Single pixel charge (gain) of an MPPC (serial number TA8120) as a function of the
overvoltage $\Delta V$ at different temperatures.}
\label{fig:gain_vs_dv}
\end{figure}

The overvoltage ($\Delta V$) is calculated by subtracting the
breakdown voltage from the operating
voltage. Fig.~\ref{fig:gain_vs_dv} shows the single avalanche charge
as a function of $\Delta V$. The fact that the curves lie on top of
each other shows that the temperature dependence of the gain is
dominated by the temperature dependence of $V_{\rm BD}$. The slopes of the
curves are consistent with the 90~fF pixel capacitance estimated from
the direct measurement, to within the equipment calibration accuracy. A
detailed analysis of the temperature dependence of the capacitance
measured at $\Delta V$=1.3~V, shows a 0.1\% increase per degree, which
can be attributed to a change in the permittivity of the silicon~\cite{krupka}.

\begin{figure}[htb]
\centering\includegraphics[width=0.4\textwidth]{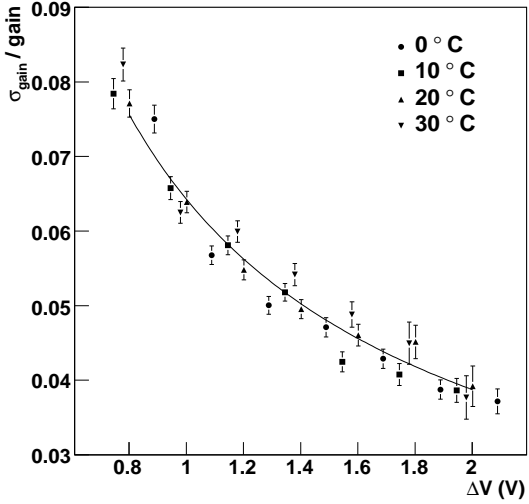}
\caption{Relative gain fluctuation vs. overvoltage at various temperatures. The
curve is a fit to the T=20$^{\circ}$C data.}
\label{fig:sigma_gain}
\end{figure}

Fig.~\ref{fig:adc_spectrum} shows that, unlike photomultiplier tubes,
the MPPC gain fluctuations are significantly smaller than the charge
from a single photoelectron avalanche. The gain fluctuations are,
however, not negligible. The spectrum presented in
Fig.~\ref{fig:adc_spectrum} can be fit by a series of Gaussian
distributions, with the $\mu$ parameter for each Gaussian representing
the mean charge in
the peak and $\sigma$ its width due to gain fluctuations and
electronics noise. The gain fluctuation
parameter $\sigma(i)$ of the $i$th peak is well described by the
equation:
\begin{equation}
\sigma(i)^{2}=\sigma_{\rm ped}^{2}+i\cdot\sigma_{\rm Gain}^{2}
\end{equation}
where $\sigma_{\rm ped}$ is the width of the pedestal, which is entirely due
to the electronics noise, and $\sigma_{\rm Gain}$ accounts for the gain
fluctuations. Measurements of $\sigma_{\rm Gain}$ show that it increases
slightly with overvoltage. However, the achievable photoelectron
resolution is related to gain fluctuation relative to the measured gain, $G$, so in Fig.~\ref{fig:sigma_gain} we show the ratio
\begin{equation}
\label{eq:mppc_sigma_gain}
\frac{\sigma_{\rm Gain}}{G}=\frac{\sqrt{\sigma(1)^{2}-\sigma_{\rm ped}^{2}}}{G}
\end{equation}
as a function of overvoltage, where $\sigma_{\rm ped}$ is the pedestal width and $\sigma(1)$ is the width of the single avalanche peak.

The 20$^{\circ}$C data can be parameterized by the following
function: $\sigma_{\rm Gain}/G=0.064\cdot\Delta V ^{-0.73}$. The quality
of the fit is good but we have no physical justification for this
particular form. There appears to be a slight temperature
variation, with the fluctuations being larger at higher temperatures.

\subsection{Dark noise}\label{dark_noise}
Dark noise in Geiger-mode avalanche photodiodes is caused mainly by
charge carriers generated thermally within the depletion region, which
then enter the Geiger multiplication area and trigger
avalanches. Any avalanche can, in turn, initiate secondary
avalanches through afterpulsing and crosstalk. Thus, the dark noise
consists of single pixel avalanche pulses, along with larger amplitude
pulses generated by optical crosstalk, afterpulsing, and accidental
pile-up from independent pixels. The last effect is negligibly small
at dark rates below 1~MHz, assuming a short integration time at the
MPPC output. Optical crosstalk and afterpulsing are discussed in the
next sections.

Since most subsystems of our experiment acquire data as charge spectra
within an integration gate associated with the beam crossing time, the
relevant dark noise metric is the fraction of these gates populated by
one or more dark pulses. The true rate of avalanches initiated by
single charge carrier can be obtained from a Poisson distribution, using
the following formula:
\begin{equation}
\label{eq:dark_rate_poisson}
R_{\rm DN}=-{\rm ln}\left(\frac{n_0}{N}\right)/\Delta t
\end{equation}
where $n_0$ stands for the number of events with no counts, $N$ for
the total number of events, and $\Delta t$ for the gate time.  The
measurements presented here used 160~ns gates triggered at
a constant rate of 20~kHz.

\begin{figure}[htb]
 \centering\includegraphics[width=.5\textwidth]{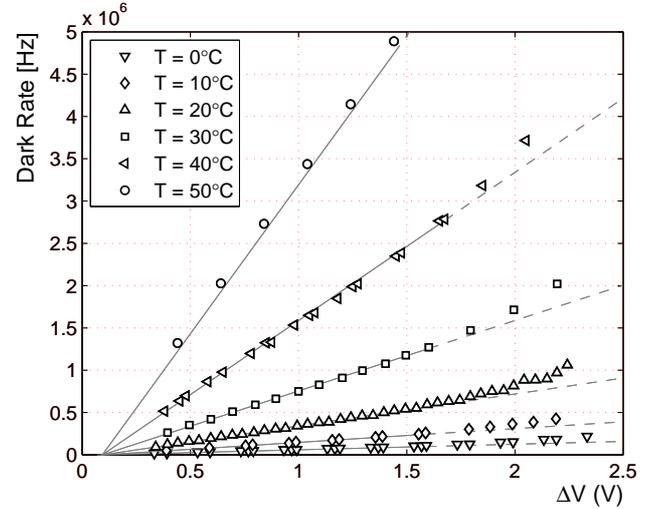}
    \caption{Dark rate vs. overvoltage $\Delta V$ at different
    temperatures (sensor number TA8120). A single fit to
    Eq.~(\ref{eq:dark_rate_fit}) has been used to fit
    all the data points. Solid lines show results within the fit range
    ($\Delta V\leq1.6$~V and $R_{\rm DN}\leq5$~MHz) while dashed lines
    represent extrapolations.}
    \label{fig:darkrate_vs_V}
\end{figure}

\begin{figure}[htb]
 \centering\includegraphics[width=.5\textwidth]{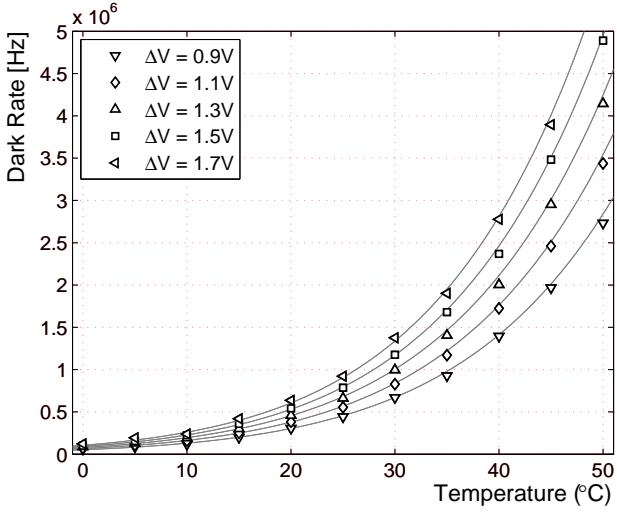}
    \caption{Dark rate vs. temperature at different
    overvoltages $\Delta V$ (sensor number TA8120). A single fit to
    Eq.~(\ref{eq:dark_rate_fit}) has been used
    to fit all the data points, with the fit range restricted to
    $\Delta V\leq1.6$~V and $R_{\rm DN}\leq5$~MHz.}
    \label{fig:darkrate_vs_T}
\end{figure}

Fig.~\ref{fig:darkrate_vs_V} shows that the dark rate increases
linearly with overvoltage in the range of 0.5--1.6~V.  Above 1.6~V the points deviate
upwards from the linear fit, which we attribute to an effect of afterpulsing.
The temperature dependence is exponential
and is shown in Fig.~\ref{fig:darkrate_vs_T}. The data for
each sensor has been fit with a
function of the form given in Eq.~(\ref{eq:dark_rate_fit}).
\begin{equation}
\label{eq:dark_rate_fit}
R_{\rm DN}\left(\Delta V,T\right)=A\cdot\left(\Delta V-V_0\right)\cdot \left( \frac{T}{298}\right)^{3/2}\cdot e^{\textstyle -\left(\frac{E}{2kT} - \frac{E}{2k\cdot298}\right)}
\end{equation}
where $T$ is absolute temperature.
In this formula $A$ represents the ratio of dark rate to overvoltage at
T=298~K (25$^{\circ}$C) (in kHz/V). $V_0$ is the offset of breakdown voltage
calculated from the dark rate with respect to that obtained from
the gain measurements, and $E$ the band gap energy. The fit range was
restricted to $\Delta V\leq 1.6$~V and $R_{\rm DN}\leq 5$~MHz, in order to
avoid the effect of afterpulsing and rate limitations of the
equipment. 

%%
%% -- Start changing text from here -- Marcin, 2011.01.08
%%

The parametrization given in Eq.~(\ref{eq:dark_rate_fit}) has been obtained under following assumptions:
\begin{enumerate}
\item A non-degenerate semiconductor model was used.
\item Thermally generated charge carriers are a result of trap-assisted (i.e. involving an R-G center\footnote{Recombination-Generation center.}) generation processes.
\item Given high reverse bias, the device operates in the so called 'R-G depletion region' steady state, i.e. no free charge carriers are available within the depletion region.
\item The trap energy level is close to the middle of the silicon's bandgap.
\item Only processes occuring within the volume of the depletion region are taken into account. Surface generation/recombination is neglected.
\end{enumerate}
Using such model, one can easily explain significant sensor-to-sensor variations of the dark rate by: a) differences in the concentrations of traps (R-G centers) and b) differences in dopant concentrations, hence different junction volumes. Mean value of the observed bandgap energy for the five measured sensors (Table ~\ref{tab:darkrate_V_fit}) is 1.127$\pm$0.0099 eV, which is within the range of values widely reported for silicon. Furthermore, reasonable $\chi^2/\nu$ values and an average \mbox{\textit{p}-value} of 33.4\% do not provide enough evidence to reject the parametrization given by Eq.~(\ref{eq:dark_rate_fit}) at a statistically significant level, which is why we assumed that it can be used to approximate data from our measurements.

%%
%% -- End changing here -- Marcin, 2011.01.08
%%

\begin{table}[htb]
	\centering
		\caption{Fit parameters for the dependence of the dark
		rate on
		$\Delta V$ and temperature.
		Eq.~(\ref{eq:dark_rate_fit}) was used with the fit range
		restricted to $\Delta V\leq1.6$~V and $R_{\rm DN}\leq5$~MHz.
        $A$ is the dark rate to overvoltage ratio at
		T=298 K (25$^{\circ}$C). $V_0$ is the difference between the
		breakdown voltage calculated from gain and dark noise.
		$E$ is the band gap energy.}
		\label{tab:darkrate_V_fit}
		\begin{tabular}{l l l l l}
			\hline
			Sensor no. & $A$ (kHz/V) & $V_0$ (mV) & $E$ (eV) & $\chi^2/\nu$ \\
			\hline
			TA8744  &	934$\pm$5.0 & 73$\pm$5.2 & 1.117$\pm$0.0024 & 1.10 \\
			TA8160  &	562$\pm$3.7 & 98$\pm$6.1 & 1.139$\pm$0.0030 & 0.99 \\		
			TA8120  &	564$\pm$3.7 & 93$\pm$6.4 & 1.135$\pm$0.0035 & 1.15 \\
			TA8092  &	622$\pm$3.8 & 92$\pm$5.9 & 1.126$\pm$0.0034 & 1.07 \\
			TA9314  &	789$\pm$3.8 & 72$\pm$4.7 & 1.118$\pm$0.0022 & 0.71 \\			
		 	\hline
		 	Mean 		&							&						 & 1.127$\pm$0.0099 &      \\
		 	\hline
		\end{tabular}
\end{table}

The dark noise rate varies significantly between MPPCs as reported
in~\cite{mppcQA}. A 20\% variation in the dark noise rate was found at
$20^{\circ}$C and $\Delta V=1$~V when calculating the variation as
the root mean square (RMS) over the mean for the 17,686 tested MPPCs.

%-----------------------------------------------------------
\subsection{Afterpulsing}\label{Afterpulses}
%-----------------------------------------------------------
%-----------------------------------------------------------
\subsubsection{Correlated noise}
%-----------------------------------------------------------
Correlated noise is a general label for avalanches that are triggered
by other avalanches. There are two known types of correlated noise:
crosstalk and afterpulsing, both of which will be described and
characterized in details in the next two sub-sections. In general,
whenever an avalanche occurs there is a chance that it triggers one or
more additional avalanches, either in neighboring pixels or in the
same pixel at a later time. As mentioned earlier, the dark noise rate
was estimated by counting the number of time no avalanches were
detected within a gate. Indeed, the average number of avalanches
detected within the gate is not a good estimator of the dark noise
rate because some avalanches may have occurred due to correlated
noise. Hence, in the presence of correlated noise, the measured
average number of avalanches will exceed the expectation from
Poisson statistics. Conversely, the measured number of events having
one avalanche within the gate will be lower than the
expectation. This fact can be used to get an estimate of the
correlated noise.

\begin{figure}[htb]
     \centering\includegraphics[width=.5\textwidth]{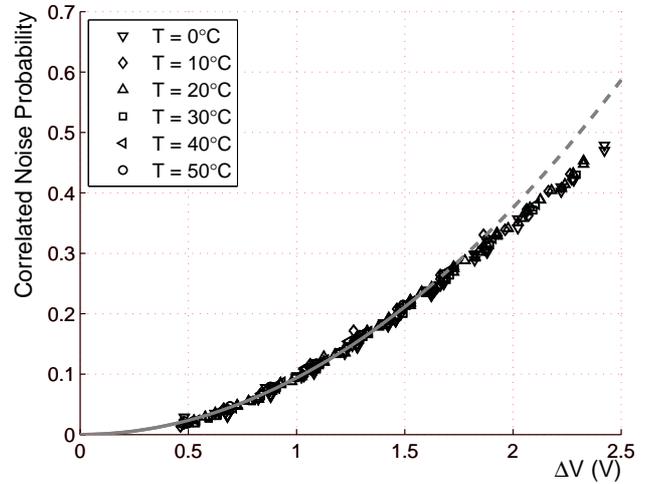}
     \caption{Combined crosstalk and afterpulse probabilities vs.
     overvoltage at several temperatures.}
     \label{fig:ct_ap}
\end{figure}

The data used for measuring dark noise presented in the previous
section can also be used to get an estimate of the correlated
noise. From the dark noise rate measurement one can predict the fraction
of events that should have one avalanche in the absence of correlated
noise. The correlated noise probability $P_{\rm CN}$ is the probability
that one avalanche triggers at least one additional avalanche. The presence of
this correlated noise term modifies the calculation of the number of events with one avalanche within the gate, $N_{1}$, as follows,
\begin{equation}
\frac{N_{1}}{N_{\rm tot}} = {e^{-R_{\rm DN}\Delta t}} {R_{\rm DN}\Delta t(1-P_{\rm CN})}
\end{equation}
where $N_{\rm tot}$ is the total number of events recorded, $R_{\rm DN}$ is
the dark noise rate, and $\Delta t$ is the gate width. The correlated noise probability
estimated by solving this equation for $P_{\rm CN}$ is shown in Fig.~\ref{fig:ct_ap}
at different overvoltages and temperatures.

The temperature dependence is strikingly small. The data can be fitted
by a quadratic function: $P_{\rm CN}=k\Delta V^2$ with k=9.4$\pm$0.1~\%. The quadratic fit is good until $\Delta V>1.6$~V, which is also approximately the overvoltage at which the measured dark noise rate no
longer behaves linearly. As explained earlier, at sufficiently large
overvoltage the method used to estimate dark noise becomes
compromised by afterpulse avalanches that stem from dark noise
avalanches prior to the integration gate. Hence, it is likely that the failure
of the fit results from the dark noise rate being incorrectly inferred
when $\Delta V$ is greater than about 1.6~V.

\subsubsection{Measuring afterpulsing}

Afterpulsing is understood as being caused by the trapping of charge
carriers created during an avalanche.
%within the depletion region.
The trapped carriers eventually get released and trigger an avalanche
within the same pixel as the original avalanche, but delayed in
time. Afterpulsing may be partially suppressed by the fact that the
pixel voltage recovers in about 45~ns (a 13.4~ns time constant) as
described in Section~\ref{recovery}. If a carrier is released while the
pixel voltage has not reached the nominal voltage, then the charge
produced in the avalanche will be lower than for nominal
avalanches. For a self-consistent description of the data, it is best
to factorize recovery and afterpulsing phenomena, i.e. to measure the
number of afterpulse avalanches per original avalanche regardless of
the pixel voltage at the time of the afterpulse avalanche. Because there
may be different types of traps in the silicon, there is no reason to
assume that afterpulsing should follow a single exponential decay. In
fact previous measurements on a similar MPPC~\cite{Afterpulsing} have
shown that afterpulsing exhibits two time constants.

Two methods were used to measure afterpulsing. The methods complement each
other since they effectively probe different afterpulsing time constants. Both
rely on the fact that afterpulse avalanches are correlated in time with their
parent avalanche.

The first method is based on the analysis of waveforms
described in~\cite{Afterpulsing}. The waveforms were fit with a
superposition
of single avalanche response functions that allow separating pulses
occurring within a few nanoseconds of one another.  The probability of an
avalanche occurring at time $t$ after another avalanche can be
parameterized as:
\begin{equation}
P(t) = \left[1 - \int_0^t P_{\rm AP}(x) dx\right]\cdot P_{\rm DN}(t) + \left[1 - \int_0^t P_{\rm DN}(x) dx\right]\cdot P_{\rm AP}(t)
\end{equation}
where $P_{\rm AP}$ and $P_{\rm DN}$ are the afterpulsing and dark noise
probabilities. [We note that the formula in Ref.~\cite{Afterpulsing} is
incorrect and should be replaced by this one.] Afterpulsing can be parameterized
using two parameters: $n_{\rm AP}$,
the number of afterpulse avalanches created per original avalanche, and
$\tau$, the time constant of the exponential distribution governing the
afterpulse generation.
A drawback of this method is that the likelihood of uncorrelated dark signals (with a rate of 500-1000~MHz) in the waveform limits the sensitivity to afterpulsing time constants of less than about 100~ns.
However, this method is well suited for measuring small time constants ($<~50$~ns) as the pulse finding techniques allow detecting pulses separated by a few nanoseconds.

The second method is based on counting the number of avalanches with a scaler after introducing a controlled deadtime following each detected avalanche.
The width of the analog pulse resulting from the convolution of the MPPC and amplifier response was such that the minimum deadtime that could be set was 26~ns. This minimum gate width means that this method is sensitive only to afterpulsing time constants of greater than about 50~ns.
However, in contrast to first method, the counting technique overcomes the dark noise background limitation when measuring long time constants by taking very high statistics data.
The deadtime dependent rate can be fit by a function that includes the contribution of
dark noise and afterpulsing. In the absence of afterpulsing, the
measured rate $R(\Delta t)$ for a given deadtime $\Delta t$ is
$R_{\rm DN} / (1+R_{\rm DN} \Delta t)$. Afterpulsing produces
avalanches that will increase the rate as long as they occur after the
deadtime.  To first order (i.e. assuming that
one avalanche creates at most one additional detectable afterpulse
avalanche and ignoring afterpulse avalanches created by previous
afterpulse avalanches) the measured rate can then be calculated from:
\begin{equation}
R(\Delta t) = R / (1 + R \Delta t)
\end{equation}
with (assuming two afterpulsing time constants)
\begin{equation}
R = R_{\rm DN} / (1- {n_{\rm AP}}_{0} e^{-\Delta t/\tau_{0}} - {n_{\rm AP}}_{1} e^{-\Delta t/\tau_{l}})
\end{equation}
where $R_{\rm DN}$ is the dark noise rate, $n_{\rm APi}$ ($i=0,1$) is the average number of afterpulse avalanches per original avalanche, and $\tau_i$ ($i=0,1$) the afterpulsing decay
time constant.

\begin{figure}[htb]
     \centering\includegraphics[width=.5\textwidth]{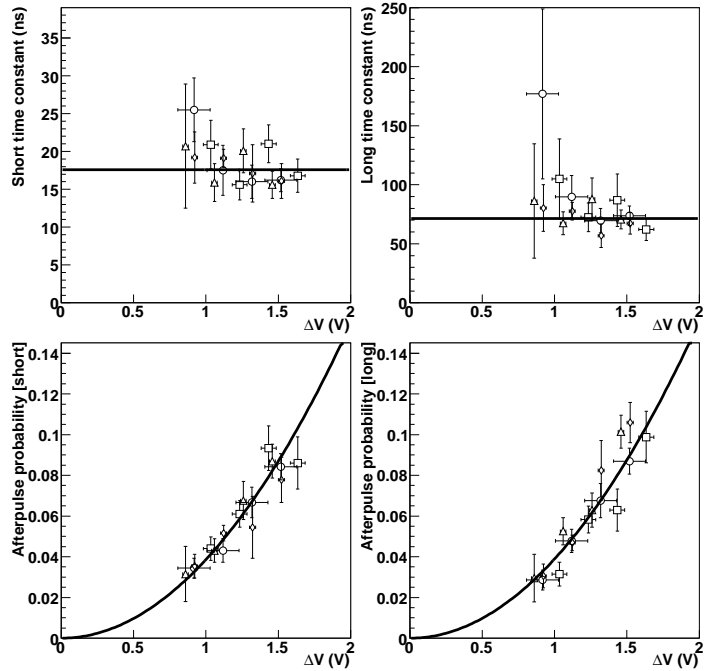}
     \caption{Afterpulse parameters (exponential time constant and probability of having at least one afterpulse) vs. overvoltage for four different MPPCs (with plot symbols: square, circle, triangle,diamond) at 25$^{\circ}$C.}
     \label{fig:Afterpulse-constants}
\end{figure}

Fig.~\ref{fig:Afterpulse-constants} shows the afterpulse
parameters for four different MPPCs measured at $25^\circ$C using the
waveform analysis technique.
%FR 11/25/10
The probability is calculated from the number of afterpulse avalanches per original avalanche distributed as $1-e^{-n_{\rm AP}}$ and so is the probability that an avalanche generates at least one afterpulse avalanche.
The exponential time constants were found to be
$\tau_{\rm short}=17.6\pm2.1$~ns and $\tau_{\rm long}=71.4\pm8.3$~ns.
The probabilities of ``long'' and ``short'' afterpulses are almost
equal.
The total probability of afterpulses is about 0.16 per
initial avalanche at $\Delta V$=1.4~V. The number of afterpulses per
avalanche as a function of overvoltage can be fit by a simple
quadratic function: $n_{\rm AP}(\Delta V)=K\cdot\Delta V^{2}$, with
$K_{\rm short}=0.0400\pm0.001(stat)\pm0.005(sys)V^{-2}$ and
$K_{\rm long}=0.0402(stat)\pm0.001\pm0.005(sys)V^{-2}$. The
dominant systematic error arises from the inability to detect pulses
less than 2~ns after the first avalanche.

\begin{figure}[htb]
     \centering\includegraphics[width=.4\textwidth]{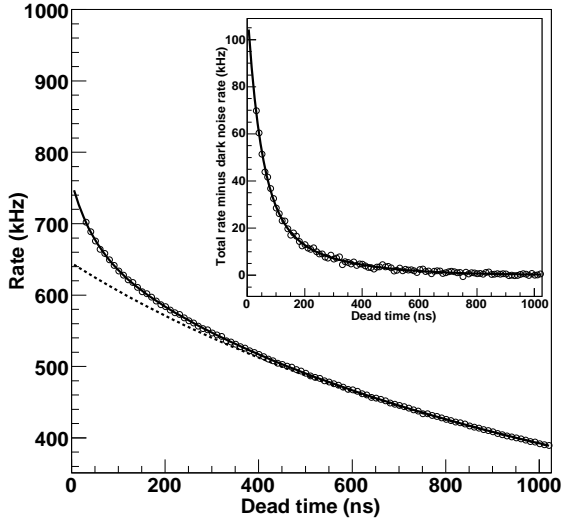}
     \caption{Rate as a function of deadtime for an MPPC biased at
     1.4~V overvoltage. The black curve shows a fit function including
     dark noise and two afterpulsing time constants.
     The dashed curve shows the estimated contribution of dark noise, i.e. the fit to Eq. (8) with afterpulsing probabilities turned off. The upper right inset shows the same data after subtracting the estimated dark noise.}

     \label{fig:ap_deadtime}
\end{figure}

Fig. ~\ref{fig:ap_deadtime} shows the rate measured
as a function of deadtime from 26~ns to 1~$\mu$s at an
overvoltage of 1.4~V at 25$^\circ$C. The fit to the data is excellent with
an average $\chi^{2}$ of 74.8
for 95 degrees of freedom. The fit parameters for the data from the
MPPC shown in Fig.~\ref{fig:ap_deadtime} yield $\tau_{0}=57\pm5$~ns,
${n_{\rm AP}}_{0}=0.107\pm0.005$, $\tau_{1}=287\pm49$~ns, and
${n_{\rm AP}}_{1}=0.043\pm0.006$. Repeating this test over 35 different
MPPCs yield the following averages: $\tau_{0}=52(8)$~ns,
${n_{\rm AP}}_{0}=0.105(0.009)$, $\tau_{1}=315(84)$~ns and
${n_{\rm AP}}_{1}=0.066(0.01)$,
with the standard deviations indicated in parentheses.

Since the two measurement methods are sensitive to different afterpulsing
time constant ranges it is not surprising they yield different
results.
It is possible to reconcile both methods by fitting the variable deadtime data for all
MPPCs with three different afterpulsing time constants, two of them being fixed: $\tau_0$=17~ns and $\tau_1$=70~ns.
The third
time constant is a free parameter in the fit. Excellent fits are again
obtained with an average $\chi^2$ of 72.52 for 95 degrees of freedom,
which is slightly better than the fit with just two time constants.
The parameters averaged over the 35 MPPCs are ${n_{\rm AP}}_{0}=0.058(0.03)$,
${n_{\rm AP}}_{1}=0.090(0.008)$, ${n_{\rm AP}}_{3}=0.056(0.009)$, and $\tau_{3}=373(55)$~ns, with the standard deviation
among MPPCs in parentheses. The expectation from the waveform analysis
at 1.4~V is ${n_{\rm AP}}_{\rm short}=0.078$ and ${n_{\rm AP}}_{\rm long}=0.082$.
The introduction of this third (373~ns) time constant into the fitting function
used for the waveform analysis at the level suggested by the variable deadtime analysis
does not worsen the agreement with the data significantly and so is an acceptable additional parameter in a range not accessible to the method.
Both analyses are also qualitatively consistent with the simple correlated noise
analysis presented in the previous section, which predicts a total contribution of
0.184 for crosstalk and afterpulsing at $\Delta V=1.4$~V.

\begin{figure}[htb]
     \centering\includegraphics[width=.5\textwidth]{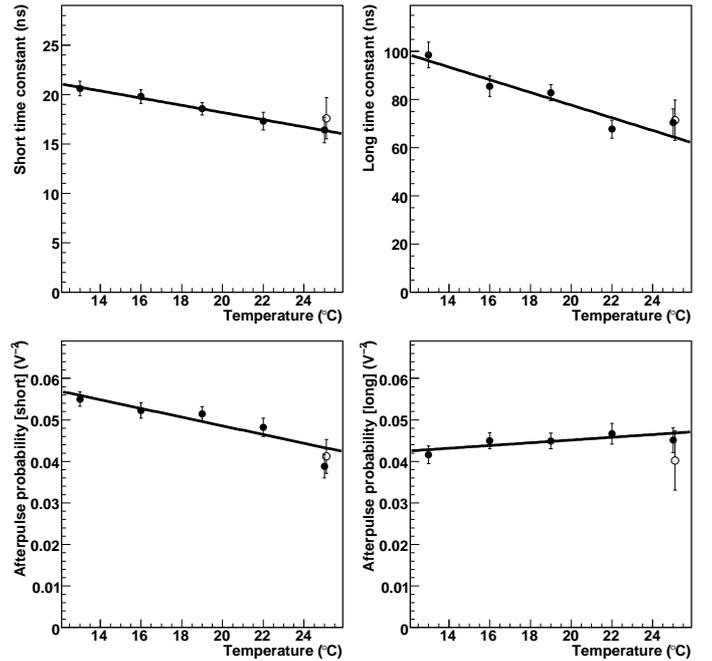}
     \caption{Afterpulse parameters (exponential time constant and probability of having at least one afterpulse) vs. temperature.}
     % The number of afterpulses per avalanche is given at an overvoltage of 1.0~V.}
     \label{fig:Afterpulse-Temp}
\end{figure}

The temperature dependence of afterpulsing was measured with the waveform
technique for a couple of MPPCs at constant $\Delta V$ within a range of 13--25$^{\circ}$C as shown in Fig ~\ref{fig:Afterpulse-Temp}. The amplitude of the ``long'' component of the
afterpulsing rate is insensitive to temperature within measurement accuracy.
On the other hand, the amplitude of the ``short'' component decreases as the
temperature increases with a coefficient of 2.0-2.5\% per
$^{\circ}$C. The short and long time constants decrease with
increasing temperature from 21~ns and 90~ns at 13$^{\circ}$C to 17~ns
and 70~ns at 25$^{\circ}$C respectively. In the MPPC simulation code,
the temperature dependence of the 17~ns and 70~ns time constants is
implemented, while the 370~ns time constant will be assumed to be
constant since no temperature dependent data are available to quantify the variation.

\subsection{Optical crosstalk}\label{Crosstalk}
%-----------------------------------------
\subsubsection{Crosstalk measurement}
%-----------------------------------------
Optical crosstalk is believed to occur when optical photons produced in an avalanche propagate to neighboring pixels where they produce photoelectrons~\cite{impactionisation}. The result is that two or more pixels can be fired almost simultaneously (i.e. on a timescale of $\approx$1~ns). The photon emission probability has been estimated to be  10$^{-5}$  photons per carrier crossing the junction~\cite{lac1}, the absorption length for photons that are most effective in propagating the avalanches is typically 1~mm.
%Trenches around pixels designed to give better isolation reduce this type of crosstalk to the level of 5-10~\% of the total signal output.
% AV : as Oleg pointed out there is no trenches. There seem to be however some metal isolation but this is not confirmed so better remove the sentence.
Although the total crosstalk fraction is small, it is expected to vary with overvoltage and a detailed study is necessary to fully characterize local variations of the crosstalk phenomenon. Measurements of crosstalk variations within the pixel array as a function of voltage were performed using optical microscopy and waveform analysis.

%----------------------------------------------------------------------------
\subsubsection{Optical microscopy}
%-----------------------------------------
\begin{figure}[htb]
     \centering\includegraphics[width=.45\textwidth]{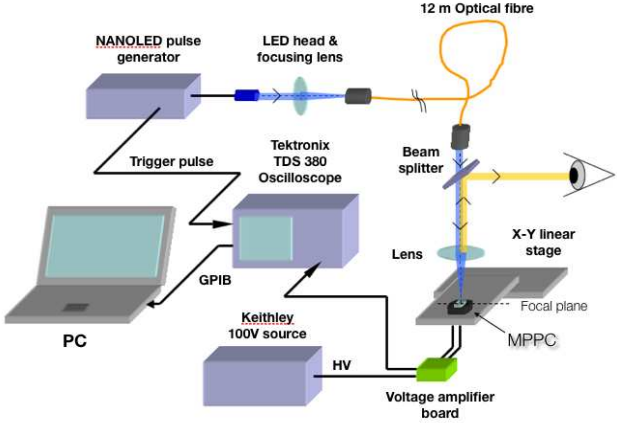}
     \caption{Schematic of the optical microscopy apparatus used to measure MPPC crosstalk within a pixel.}
     \label{fig:ctsetup}
\end{figure}
Crosstalk probabilities were measured using the apparatus shown schematically in Fig.~\ref{fig:ctsetup}. A {\it nanoLED}~\cite{nanoLED} light source system was used to produce a pulse width of 1~ns FWHM from an integrated 463~nm LED.
This was coupled to an optical fiber that terminated in a microscope lens such that the light beam is focused onto the MPPC face. The MPPC was mounted on an X-Y stage so that the beam spot could be translated across the MPPC pixel array with one micron position resolution. The MPPC signal was digitized with a 1~GHz sampling rate during a 1~$\mu$s period using a Tektronix TDS~380 oscilloscope. The light pulse intensity was measured to be between 10--20 photons at the exit point of the fiber.

\begin{figure}[htb]
 	 \centering\includegraphics[width=.5\textwidth]{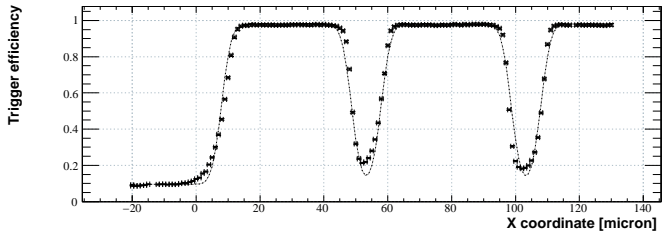}
     \caption{MPPC detection efficiency across several MPPC pixels. The dashed line is the expected profile from the convolution of a Gaussian beam spot with $\sigma=5~\mu$m  and square 50~$\mu$m pixels.}
     \label{fig:wfandprofile}
\end{figure}

It was assumed that the amplitude of the avalanche signal observed is not dependent on the number of photons injected into a pixel if the photons all originate from the same LED pulse.  Therefore each LED flash creates a 1~p.e. signal as long as the beam spot is well-centered within a pixel. The trigger efficiency was measured as the beam was scanned across several pixels to estimate the profile of the photon beam. The profile measured over 150~$\mu$m is shown in Fig.~\ref{fig:wfandprofile} and is consistent with a Gaussian beam spread of 5~$\mu$m. The sensitive area is in good agreement with the value of 61.5\% specified in the Hamamatsu catalog.

%----------------------------------------------------------------------------
\subsubsection{Crosstalk study using waveform analysis}

\begin{figure}[hbt]
\centering
\includegraphics[width=.5\textwidth]{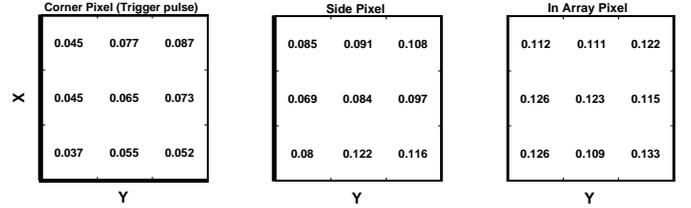}
\caption{Measured crosstalk probabilities for nine beam positions inside an MPPC pixel at $\Delta V=1.335$~V.  (Left) a corner pixel with the corner located on the bottom left; (center) a side pixel with the side boundary to the left of the pixel and (right) a pixel inside the MPPC array.  The thick black line denotes the limit of the pixel array.}
\label{fig:ctvspos_t}
\end{figure}

Waveforms were recorded for nine beam positions inside each of three pixels chosen for their specific position within the array, namely: a corner pixel, a side pixel and a pixel inside the MPPC array away from the edges.  These pixels are surrounded by 3, 5 and 8 pixels respectively, each of which may generate crosstalk signals. Crosstalk probabilities were calculated for individual photoelectron pulses selected to be within 8~ns after the LED trigger pulse. The total crosstalk probability is given by:
\begin{equation}
P_{\rm {ct}} = 1 - \frac{N(1\text{pe})}{N_{\rm {tot}}},
\end{equation}
where N(1pe) is the number of single 1~p.e. pulses and $N_{\rm {tot}}$ is the number of all LED pulses.
The total crosstalk signal is defined as the observation of $\geq$2~p.e. pulses within the 8~ns time window, while
individual crosstalk probabilities were extracted by selecting pulse heights corresponding to 2~p.e., 3~p.e. and  4~p.e.
Data was taken for  overvoltage $\Delta V=1.335~V$ and $T=24^{\circ}$C; results for all three pixels are presented in Fig.~\ref{fig:ctvspos_t}.
For all three pixels the crosstalk measured shows a clear dependence with position of the beam spot, suggesting that the crosstalk probability is
correlated with where the photon is absorbed in the pixel.
 A similar analysis was applied to MPPC dark count data in a time window 500~ns before the LED triggers. The crosstalk was measured to be $9\pm1$\% and no correlation with beam spot location was found.

\begin{figure}[htb]
    \centering\includegraphics[width=.5\textwidth]{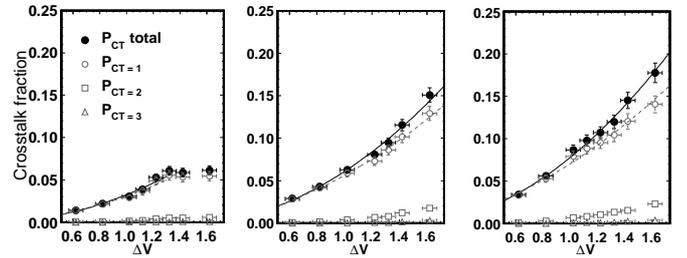}
    \caption{Crosstalk value vs. overvoltage for three pixels shown in Fig.~\ref{fig:ctvspos_t}. $P_{\rm {ct}}$ is crosstalk probability, $P_{\rm {ct=1}}$ is the probability of only one pixel fired in addition to the initial pixel, etc.}
 \label{fig:ctvsoverv}
\end{figure}

\begin{figure}[htb]
    \centering\includegraphics[width=.35\textwidth]{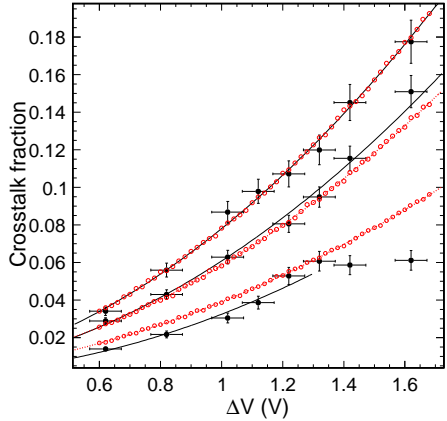}
     \caption{Total crosstalk probabilities for all three pixels (corner, side and in-array). Data is shown as a solid circle, a solid black line indicates the best quadratic fit from Fig.~\ref{fig:ctvsoverv}. Simulated probabilities are shown as open circles and a dashed line.}
 \label{fig:ctmodel}
\end{figure}

Fig.~\ref{fig:ctvsoverv} presents measurements of crosstalk probabilities as a function of overvoltage at $T=24^{\circ}$C for the same three pixels.
Based on the position variation results, a correction factor is applied to correct the crosstalk probability to a probability averaged over the entire pixel.
All probabilities were found to agree with a $\Delta V^2$ dependence except for the corner pixels where the total probability plateaus at high overvoltage (above 1.3~V). This plateau is due to some peculiar behaviors of the corner pixels, which cannot be explained by geometrical considerations. Variations of the total crosstalk probability between pixels is in good agreement with the hypothesis of a point source generation of optical photons in the pixel. The result of the crosstalk simulation is shown in Fig.~\ref{fig:ctmodel} and agrees well with the data for a nearest neighbor crosstalk hypothesis. This model is included in the simulation described in Section~\ref{Simulation}.

\subsection{Photon detection efficiency}

The  photon detection efficiency (PDE) of a multi-pixel  avalanche  photodiode
operated in a limited Geiger mode is a product of three factors:
\begin{equation}
{\rm PDE} = QE(\lambda)\cdot\varepsilon_{\rm {Geiger}}\cdot\varepsilon_{\rm {pixel}},
\label{eq:pde}
\end{equation}
where $QE(\lambda)$ is the wavelength dependent quantum efficiency, $\varepsilon_{\rm {Geiger}}$ is the probability to initiate the Geiger discharge by a carrier, and geometric acceptance $\varepsilon_{\rm {pixel}}$ is the fraction of the total photodiode area occupied by the photosensitive area of the pixels.

For an MPPC, quantum efficiency can be defined as the probability for an incident photon to generate an electron-hole pair in a region in which carriers can produce an avalanche.  The layer structure in an MPPC is optimized to have the highest probability for a visible photon to be absorbed in the depletion layer. %Since the first p$^+$ layer is about 1~$\mu$m thick the MPPC has a peak sensitivity for blue photons, which have an absorption mean free path of 0.48~$\mu$m at 450~nm. Photoelectrons produced in the p$^+$ layer drift to the high-field junction where the Geiger avalanches may occur. Since the mean free path for green (520~nm) photons is 1.36~$\mu$m, carriers from this source are produced in the n$^+$ layer behind the junction so avalanches are initiated with lower efficiency. [A discussion of the MPPC layer structure may be found in Ref. \cite{murase}.] 
For comparison, an APD with a similar layer structure to that of the MPPC developed by Hamamatsu for the CERN CMS experiment~\cite{apdqe} has a measured quantum efficiency of more than 80\% at 500~nm, so a similar value may be expected for the MPPC.

Overvoltage affects just one parameter in the expression, namely $\varepsilon_{\rm {Geiger}}$.
The breakdown probability depends  on the impact ionization coefficients (for electrons and holes), which are strong functions of electric field. Simulation and measurements~\cite{pospd09} show that $\varepsilon_{\rm {Geiger}}$ behaves as the exponentially saturating function $\varepsilon_{\rm {max}}(1-e^{-k\Delta V})$ if breakdown is triggered by electrons. Breakdown initiated by holes leads to a linear dependence on $\Delta V$.

The geometrical factor $\varepsilon_{\rm {pixel}}$ is solely determined by the MPPC topology. Our measurements indicate $\varepsilon_{\rm {pixel}}$=64\%, which is consistent with the Hamamatsu specification of 62\% for sensors with 50~$\mu$m pixels.

\subsubsection{PDE measurement--pulsed LED method}\label{pulsed}

For the PDE measurements we used an approach discussed in~\cite{musienko}. The PDE is measured using pulsed LED light with a narrow emission spectrum.  The number of photons per LED flash is collimated to be within the MPPC sensitive area and reduced to an intensity that can fire only 2-5 pixels on average.  The number of photons per LED pulse $N_{\gamma}(\lambda)$ can be measured using a calibrated photodetector, i.e. one with known spectral and single electron responses.

The PDE can be calculated from the recorded MPPC pulse height distribution (see Fig.~\ref{fig:adc_spectrum}) by assuming a Poisson distribution of the number of photons in an LED pulse. The mean value ${\bar N}_{\rm {pe}}$ of the number of photons recorded per LED pulse can be determined from the probability P(0) of the pedestal (0~p.e.) events by  ${\bar N}_{\rm {pe}}=-{\rm ln}P(0)$.  ${\bar N}_{\rm {pe}}$ calculated this way is independent of afterpulsing and crosstalk.  Knowing the number of photons incident on the MPPC, $N_{\gamma}(\lambda)$,  one can calculate ${PDE}(\lambda)={\bar N}_{\rm {pe}}/N_{\rm{\gamma}}(\lambda)$.

The dependence of PDE on bias voltage was measured using a fast green emitting LED  operating in a pulsed mode. The emission spectrum of this LED was measured to be very close to that of a Y11 WLS fiber. The peak value is centered around 515~nm, and it has FWHM of 40~nm.

\begin{figure}[htb]
\centering\includegraphics[width=0.5\textwidth]{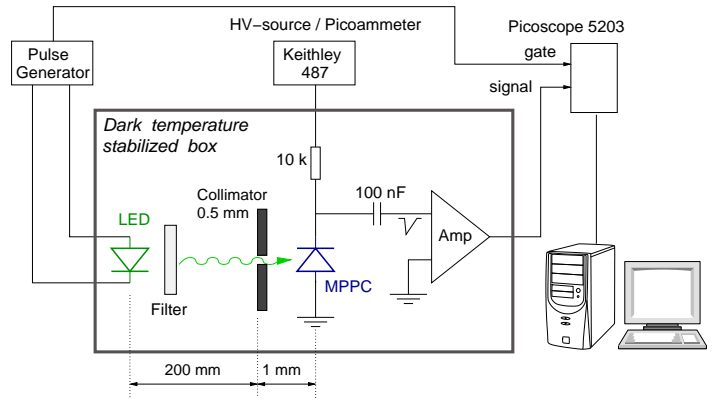}
\caption{Schematic diagram of the setup for the PDE measurements.}
\label{fig:setup}
\end{figure}

The MPPC was illuminated with LED flashes through a 0.5~mm collimator
(the distance between the LED and the collimator was 20~cm, and with about 1~mm between the
collimator and the MPPC). A neutral density filter reduces the light intensity on the MPPC face to the level of 10--15 photons. The signal from the MPPC was amplified with a fast transimpedance  amplifier  and digitized with a Picoscope 5203 digital oscilloscope (250~MHz bandwidth, 1~GHz sampling rate). The signal integration time was 150~ns. A schematic diagram of the setup  is shown in Fig.~\ref{fig:setup}.

The PDE measurements were done in a temperature stabilized dark box ($\Delta {\rm T}$$<$0.1$^{\circ}$C). The number of photons in the LED flash
was measured using a calibrated Hamamatsu photomultiplier R7899 (QE=15.7\% at 515~nm).  The photoelectron collection efficiency for the 5-mm diameter central part of the photocathode is more than 95\%, the PMT excess noise factor is 1.15.  The LED amplitude spectrum measured for one of the tested MPPCs is shown in Fig.~\ref{fig:adc_spectrum} at $\Delta V$=1.5~V and 20$^{\circ}$C.

The average number of photoelectrons in the LED signal was calculated by counting the number of pedestal events as discussed in Section~\ref{dark_noise}. To correct for dark pulses that occurred randomly inside the 150~ns integration gate, the dark rate was measured during the same gate width but 300~ns earlier relative to the LED pulse.  The stability of the LED pulse intensity  was monitored and found to be better than $\pm 3$\% during  the measurements.

\begin{figure}[htb]
\centering\includegraphics[width=0.45\textwidth]{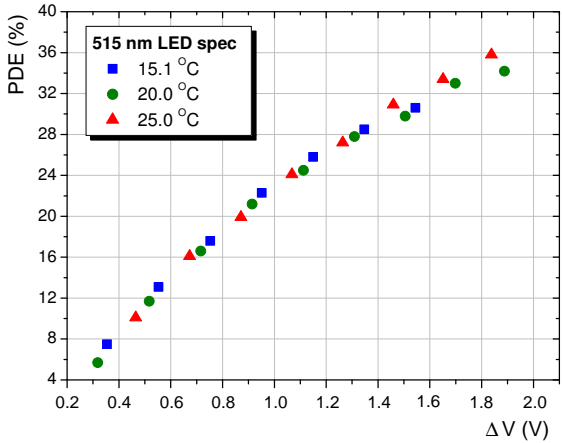}
\caption{ Photon detection efficiency of a MPPC (serial number TA9445) for green light ($\sim$515~nm) as a function of overvoltage at three temperatures.}
\label{fig:pde_3temperature}
\end{figure}

MPPC PDE as a function of overvoltage is shown in Fig.~\ref{fig:pde_3temperature} for three temperatures. The PDE depends almost linearly on $\Delta V$ within the $\Delta V$ range of 1.0--1.6~V with slope of 1.5\% per 100~mV. For a fixed overvoltage there is no observable dependence on temperature.

\begin{figure}[htb]
\centering\includegraphics[width=0.45\textwidth]{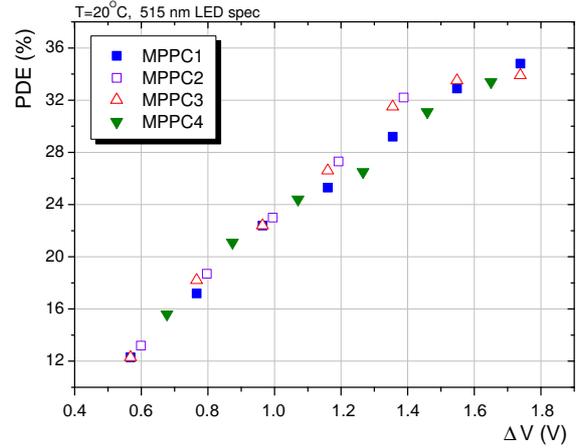}
\caption{PDE (at 515~nm) vs. overvoltage for four MPPCs  at  20$^{\circ}$C.}
\label{fig:pde_4overVs}
\end{figure}

Fig.~\ref{fig:pde_4overVs} shows the measured PDE for four additional MPPCs at 20$^{\circ}$C. All show essentially the same performance with the PDE in the range 29--32\% for green light at a typical operating overvoltage of 1.4~V.
The measurement accuracy of the PDE is estimated to be about 10\%. The largest contribution to this uncertainty is the normalization error, which is dominated by the error in the PMT spectral calibration (5-7\%) followed by the uncertainty in the p.e. collection efficiency in the PMT (5\%).

\subsubsection{PDE measurement--optical power meter method}
The PDE has also been measured  independently using a 473~nm LED pulser developed as a calibration source for the ANTARES experiment~\cite{antares_led} and with a 463~nm  NanoLED pulser. The experimental setup is similar to that shown in Fig.~\ref{fig:setup}. The MPPC signal was amplified with a gain of 40 and then sampled by a LeCroy WaveRunner 6100 oscilloscope (1~GHz bandwidth, 10~GSample per sec) within a 200~ns gate.
The temperature was held stable to 0.1$^{\circ}$C by means of a thermally-coupled metallic plate.

Calibration of the number of photons incident on the MPPC was performed using
a Newport 1835-C Optical power meter with an 818 series 1~cm$^2$ photodiode
sensor. The power meter converts an optical signal of specific wavelength into the optical power equivalent. The number of photons per flash can be calculated as follows,
\begin{equation}
N_\gamma =  \frac{P_{\rm {W}}\cdot\lambda}{F_{\rm {Hz}}\cdot hc}\cdot A,
\end{equation}
where $P_\text{W}$ is the measured optical power (in watts) at wavelength $\lambda$ (463 or 473~nm), $F_{\rm {Hz}}$ is the LED pulse rate (13~kHz), $h$ is Planck's constant and $A$ is an acceptance factor.
Acceptance $A$ is the ratio of intensity of LED light incident on the MPPC sensitive area to the intensity incident on the power meter sensor. The value of $A$ is evaluated by performing a position scan of the LED light intensity profile.

The power meter calibration accuracy and the estimated acceptance factor are the dominant contributors to the uncertainties in the PDE measurement. As discussed in Section~\ref{pulsed}, the number of detected photoelectrons was obtained from the number of pedestal events in signal and assuming a Poisson distribution. The PDE values are free of afterpulsing and crosstalk contributions and were corrected for dark rate. The PDE was found to be 31\% for the 463~nm LED and 29\% for the 473~nm LED at $\Delta V$=1.3~V, which is in good agreement with the MPPC spectral sensitivity shown in Fig.~\ref{fig:spec} and discussed in the next section.

\subsubsection{Spectral sensitivity}
A spectrophotometer calibrated with a PIN-diode~\cite{musienko} was used to measure the spectral sensitivity of the MPPC. The spectrophotometer light intensity was reduced until the maximum MPPC current was only $\sim$30\%  greater than the dark current to avoid nonlinearity effects caused by the limited number of pixels. Comparing the MPPC current  with the calibrated PIN-diode photocurrent we obtain the relative spectral sensitivity.

\begin{figure}[htb]
\centering\includegraphics[width=0.5\textwidth]{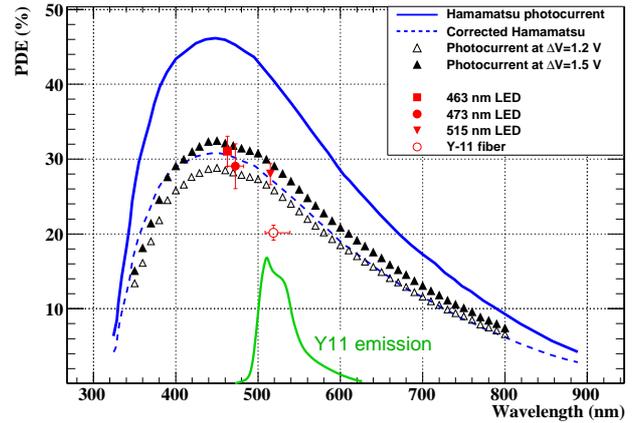}
\caption{MPPC PDE as a function of wavelength at  $\Delta V$=1.2 and 1.5~V at 25$^{\circ}$C. Also shown is the spectral plot from the Hamamatsu catalogue, which uses data not corrected for crosstalk and afterpulsing (blue line); the dashed line is the Hamamatsu plot scaled-down using knowledge of the correlated noise contribution from our measurements.
The green curve shows the Y11(150) Kuraray fiber emission spectrum (arbitrary units) for a fiber length of 150~cm (from Kuraray spec). LED and Y11 fiber points were measured at $\Delta V$=1.3~V.}
\label{fig:spec}
\end{figure}

To achieve an absolute scale, the measured relative spectral response is scaled to the reference PDE points obtained with LED light at three wavelengths: 410, 460 and 515~nm measured at 1.2~V overvoltage. The scaling factor at other overvoltage values was found to be constant at these wavelengths up to about 1.4~V---the PDE spectral sensitivity shape is appreciably different above this, as was noted in \cite{pospd09}. The MPPC PDE dependence on the wavelength of the detected light along with the emission spectrum of the WLS Y11 Kuraray fiber are  shown in Fig.~\ref{fig:spec}.  The MPPC peak sensitivity is in the blue light region, around 450~nm.

Since the spectrum of light incident on the MPPC in the ND280 detectors is determined by the Y11 fiber emission spectrum and the wavelength-dependent attenuation in the fiber, a PDE measurement was performed by exciting a Y11 fiber with a 405~nm LED~\cite{wardPDE}.
%and calibrating out the light coming out of the fiber
The blue light source was arranged so that only re-emitted green light could reach the photosensor after propagating through 40~cm of the fiber. The fiber was coupled directly to the MPPC with the same design of optical connector used in the ND280 ECAL and P0D subsystems.
%which uses a soft sponge to apply the small pressure on MPPC towards the fiber end.
At $\Delta V$=1.3~V, the PDE was measured to be 21\%, which is significantly lower than the 28\% measured at the same overvoltage with light incident directly onto the MPPC. The lower value may be due to light loss at the interface between the coupler and the Y11 fiber.

Fig.~\ref{fig:spec} also shows the MPPC spectral sensitivity measured by Hamamatsu for a commercial MPPC S10362-11-050 device at 25$^{\circ}$C.
These data, taken from the Hamamatsu catalog, are not corrected for crosstalk and afterpulsing.
The method Hamamatsu used is basically the photocurrent method described above but with a monochromator to select the incident light wavelength. The number of incident photons is derived from a calibrated photodiode response and the number of detected photoelectrons is obtained by dividing the MPPC current by its gain and the charge on an electron and assuming a Poisson distribution of the number of photons per single flash.
We have corrected the Hamamatsu result by scaling down the PDE values by 0.663. This scaling factor was chosen to fit the sensitivity curve at the points measured with LEDs; the renormalized Hamamatsu spectral plot shape is consistent with our results within measurement accuracy.

\section{MPPC simulation}\label{Simulation}
A Monte Carlo simulation of the MPPC, written in C++, is now at a mature stage
of development. The simulation can be split into two main components - a set
of models defining device behavior, and a procedural framework to initialize
the model using input parameters, control the simulation and output the
results. The framework will be discussed briefly first.

\subsection{Simulation framework}

The simulation is based on a list of potential triggers (incident photons,
thermally generated carriers and crosstalk/afterpulses), which are processed
in time order. The only state variables of the MPPC are the voltages across
each pixel; the evolution of these voltages between triggers is handled by a
recovery model. On initialization, a list of incident photons is given to the
simulation as input, and thermal noise is generated at the appropriate rate
DCR($V_{\rm {nom}}$) for a nominal operating bias voltage and temperature.
These two sources form the initial list of potential triggers.

Each potential trigger is then processed in the following steps:

\begin{enumerate}
\item The voltages on all pixels are updated from their state after the
  previous trigger, using the recovery model and the elapsed time since the
  last processed trigger.
\item It is determined whether the pixel fires. The probability is equal to
  PDE($V_{\rm {pix}}$) for photons and
  DCR($V_{\rm {pix}}$)/DCR($V_{\rm {nom}}$) for dark noise, to account for
  the lower DCR for a pixel with depleted voltage, relative to the nominal DCR
  used to generate the noise triggers.
  %Afterpulse/crosstalk always fires since these probabilities are accounted for when the trigger was generated.
\item If the pixel fires, a trigger is added to the list of output signals and
  its voltage is set to zero; the charge of the generated avalanche depends on
  the voltage of the fired pixel and it is smeared by a Gaussian resolution
  function. The afterpulse/crosstalk models determine whether further noise is
  generated, and, if applicable, the additional noise is inserted into the list of potential hits, in correct time order.
\end{enumerate}

The reinsertion of correlated noise resulting from an initial trigger allows
higher-order noise cascades to be dealt with in a simple and natural way. The
final output is a list of avalanches with times and charges, which can then be
processed by code appropriate to a specific readout circuit.

\subsection{Physics models}
The simulation relies on accurate models for the various effects present in
the sensor. The characterization measurements described above have been used
to determine appropriate models to use, and to tune model parameters.

The dark rate is parameterized as a linear function of bias voltage--the
parameters for this function must be measured separately for each sensor since
large variations between devices are observed. The PDE is modeled with a
quadratic fit to the data in Fig.~\ref{fig:pde_3temperature}; variation with
wavelength is not included. The mean number of short- and long-lived trapped
carriers for afterpulsing, and the lifetime of the trapped states, are taken
from the results of the waveform analysis method in Section~\ref{Afterpulses}.
The crosstalk model is based on the data and the model described in
Section~\ref{Crosstalk}.
The data shown in Fig.~\ref{fig:ctmodel} are well-described by a simple nearest-neighbor model that assumes crosstalk  occurs only in the four nearest-neighbor pixels to the primary pixel. Crosstalk pulses are generated according to the probability measured from dark noise. The location of the crosstalk pulse is then chosen randomly among the four neighboring pixels. The pulse is discarded if it falls outside the MPPC active area.

The recovery model used is specific to the ND280 Trip-t-based electronics (TFB
board~\cite{tript}), it assumes recharging of the fired MPPC pixels from
capacitances elsewhere in the readout electronics for each channel. Recovery does not
significantly affect response at low light levels however, so it will not be
discussed in further detail.

\subsection{Comparison with data}

The simulation output has been compared to data taken using the ND280 Trip-t
electronics and a fast-pulsed LED, with a gate length of 540~ns and the photosensor at a temperature of 22$^\circ$C. An adjustable lens was used to alter the
intensity of light incident on the sensor. All the parameters used for the
simulation were taken from the characterization measurements, but some
tuning was required to reflect sensor-specific parameters, electronics effects
and light-level uncertainties. The linear fit parameters for the sensor dark
noise curve were measured and used in the simulation. Since an absolute
calibration of the incident light level was not available, the mean incident photon
number was calculated for 1.33~V overvoltage using the method described in Section~\ref{pulsed}. The absolute PDE in the simulation is therefore not tested by
this comparison, but errors in the parametrization of PDE as a function of
voltage will be evident. Finally, the spread in total event charge due to
electronics noise, and the spread in avalanche gains, were determined from the
measured peak widths at a low light level and 1.33~V overvoltage, and
added to the simulation.

\begin{figure}
\includegraphics[width=0.45\textwidth]{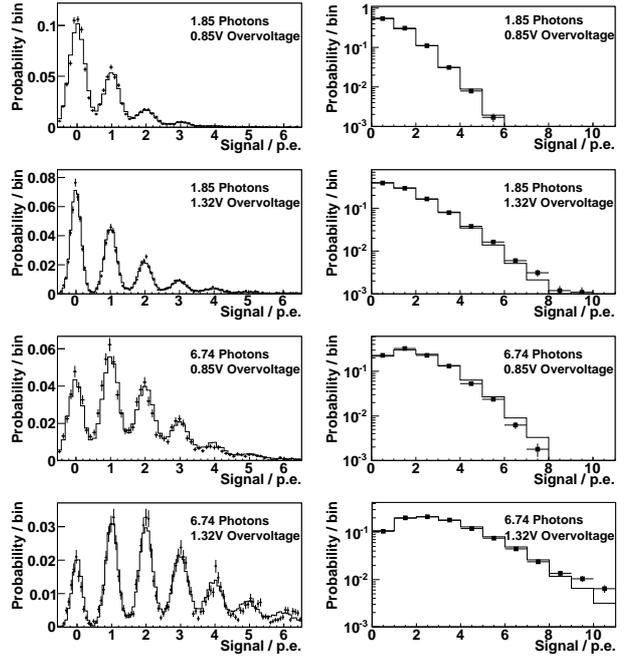}
%\caption{Comparison of data to Monte Carlo at low light level for a range of overvoltages.}
\caption{Comparison of data to Monte Carlo at low light level for a range of overvoltages. The photon numbers shown are the number incident on the face of the MPPC. The histograms on the right show the same data as on the left, but with a bin width equal to the fitted peak separation in the data.}
\label{simLedSpectra}
\end{figure}

Histograms of integrated output charge are shown in Fig.~\ref{simLedSpectra}, for data and simulation. Very good agreement is seen
between the data and MC for a range of light levels and overvoltages. Some
small discrepancies between data and MC are seen in the integer-binned histograms; however these
histograms depend on the exact peak positions, which must be determined in the
data by fitting. They also depend sensitively on the exact shapes of the
peaks, since for large peak widths, each integer bin contains some events
which have migrated from neighboring peaks. No significant systematic
difference is observed between data and MC.

\subsection{Energy resolution}

In most cases, the energy resolution of scintillator detectors is dominated by the photon counting statistics when the number of photoelectrons is low (less than about 100). However, the photosensor and electronics can impact the energy resolution in two ways: 1) constant noise background due to dark noise and electronics noise, 2) fluctuation in the charge detected per photoelectron. The energy resolution can be calculated fairly accurately in the case where the MPPC charge is integrated over a time window $\Delta t$ and ignoring the MPPC saturation effect. The standard deviation of the number of avalanches can be written as:

\begin{equation}
\sigma_{N_{\rm Av}}^2 = N_{\rm Av} + N_{\rm Av} \sigma_{\rm G}^2 + \sigma_{\rm el}^2 + R_{\rm DN} \Delta t
\end{equation}

where $N_{\rm Av}$ is the number of pixel avalanches, $\sigma_{\rm G}$ is the gain fluctuation parameter, $\sigma_{\rm el}$ is the electronics noise integrated over $\Delta t$, and $R_{\rm DN}$ is the dark noise rate.
$N_{\rm Av}$ is related to the number of photoelectrons at low light level by $N_{\rm Av}=N_{\rm PE}(1+N_{\rm CN})$, where $N_{\rm CN}$ the number of correlated noise avalanches per avalanche. This latter formula is an approximation as it does not account for gain recovery and correlated noise avalanche created by other correlated noise avalanches.
However, the MC simulation includes both effects. Some conclusions can be drawn from this formula: 1) the integration gate ($\Delta t$) should be chosen so that $ N_{\rm Av}  \gg R_{\rm DN} \Delta t$ in order to ensure that dark noise does not contribute to the resolution, 2) the gain fluctuations do not contribute to the resolution significantly since $\sigma_{\rm G}$ is only about 10\%. This last conclusion highlights a significant difference between MPPCs and PMTs or standard Avalanche Photodiodes (APDs), whose main contribution to the energy resolution arises from gain fluctuations.

\begin{figure}[htb]
\centering\includegraphics[width=0.3\textwidth]{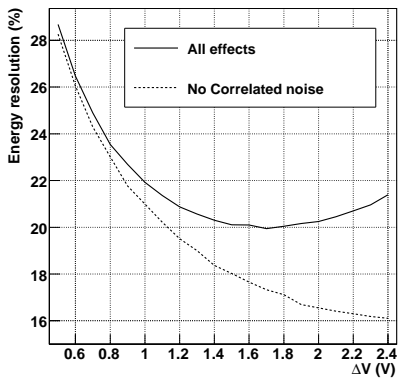}
\caption{Simulated energy resolution as a function of overvoltage for a typical MIP signal of about 25 avalanches. The confounding effects include crosstalk, dark noise and afterpulses. The curve without these effects includes only the variation of the MPPC efficiency with overvoltage.} \label{fig:energyRes} \end{figure}

The simulated energy resolution is shown in Fig.~\ref{fig:energyRes} as a function of overvoltage with and without correlated noise (crosstalk and afterpulse). A gate of 540~ns was used to integrate the charge.
The light flash occurred 60~ns after the beginning of the gate and the photons were produced according to an exponential with a 7~ns time constant. The number of incident photons was set to 100 to match the average number of avalanches triggered by a minimum ionizing particle in T2K near detectors, which ranges between 20 and 35 avalanches. Without correlated noise the energy resolution would improve with increasing $\Delta V$ because of the increasing photodetection efficiency. In practice, when correlated noise is included the energy resolution reaches a minimum at $\Delta V=1.8$~V.
Beyond 1.8~V, correlated-noise-induced fluctuations worsen the energy resolution. Due to dynamic range constraints, in the T2K ND280 the MPPCs are operated at no more than
$\Delta V=1.33$~V.

\begin{figure}[htb]
\centering\includegraphics[width=0.3\textwidth]{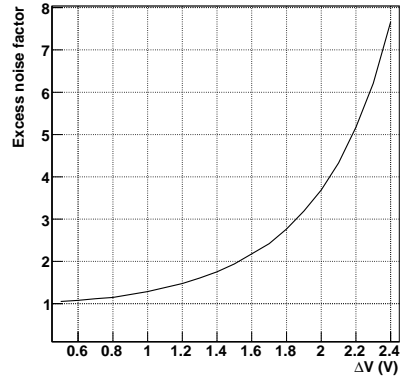}
\caption{Excess Noise Factor as a function of overvoltage.} \label{fig:enf}
\end{figure}

The detector energy resolution is dominated by the photon counting statistics when $\Delta V$ is less than about 1.5~V; above 1.5~V correlated noise contributes significantly. For photomultiplier tubes and APDs, the contributions of gain fluctuations and correlated noise to the energy resolution are often assessed by calculating the excess noise factor (ENF). This better quantifies the contribution of the photosensor and the electronics system to the resolution by dividing out the fluctuations introduced by the photon counting statistics:
\begin{equation}
F = \sigma_{N_{\rm Av}}^2/N_{\rm PE}
\end{equation}
The dependence of the excess noise factor with overvoltage is shown in Fig.~\ref{fig:enf}. The ENF increases with increasing overvoltage following the increase of crosstalk and afterpulsing, which add additional avalanches in a stochastic manner. The ENF reaches 2 at a value of $\Delta V$ of about 1.5~V. The MPPC ENF is nevertheless significantly smaller than for APDs, whose ENF is always larger than 2~\cite{APDENF}. Overall, the MPPC contribution to the energy resolution is small for minimum ionizing particles that typically yield between 20 and 40 avalanches on average, even for T2K sub-detectors operating at
$\Delta V=1.33$~V.

\section{Conclusion}

The T2K experiment ND280 complex of detectors uses a 667-pixel MPPC developed by Hamamatsu Photonics specifically for this experiment.  It has a sensitive area of 1.3$\times$1.3~mm$^2$ and a pixel size of 50$\times$50~$\mu$m$^2$; the sensitive area is larger than those available previously and relaxes the mechanical tolerances required for coupling to the WLS fibers used extensively in the experiment. We have performed detailed investigation of these devices and have developed
an accurate model of the MPPC response to low light levels (where saturation effects can be neglected).

MPPCs biased at the recommended Hamamatsu overvoltage (1.33~V) at T=25$^{\circ}$C are characterized by the following parameters: photodetection efficiency of about 20\% when illuminated with light from Y11 fibers (peaked in the green); a typical gain of 7.5$\times$10$^5$; the average dark rate is 700~kHz but can approach 1~MHz; the crosstalk and afterpulse probability are estimated to be 9-12\% and 14-16\% respectively, with a combined total of 20--25\%; and the recovery time constant of a single pixel is 13.4~ns.
With such parameters, the device achieve the desired 20\% energy resolution for
%With such parameters, the device performance does not affect the energy resolution
for minimum ionizing particles, which yield on average between 20 and 40 avalanches in the various components of the T2K near detector. Furthermore, about 40,000 MPPCs were operated in the T2K neutrino beam in 2009-10 and no significant reliability issues were experienced.

Modeling the MPPC response by parameterizing dark noise, afterpulses, photodetection efficiency, crosstalk and gain variation enables us to account for the contribution of the photosensor to the overall detector response accurately.
The MPPC saturation effect should also be fully described by our simulations, but confirmation of this awaits additional controlled measurements for final validation.

This work was supported in part by the ``Neutrino Physics'' program of the
Russian Academy of Sciences, the RFBR (Russia)/JSPS (Japan) grant \#08-02-91206, Polish Ministry of Science and Higher Education, grant number 35/N-T2K/2007/0, US Department of Energy grants DE-FG02-93ER40788 and DE-FG02-91ER40617, National Sciences and Engineering Research Council of Canada, and the UK Science and Technology Funding Council (STFC).

%% Bibiliography for the nd280 mppc NIM paper
%% bib.tex
%% Bob Wilson 25mar09

\end{document}